\documentclass[12pt]{article}
\usepackage{amssymb}
\usepackage{epsfig}
\usepackage{graphicx}
\usepackage{amsmath}
\usepackage{verbatim}
\usepackage[numbers,sort&compress]{natbib}

\csname @addtoreset\endcsname{equation}{section}
\begin{document}
\begin{titlepage}
\title{\bf\Large MSSM with $m_{h}=125$ GeV in High-Scale Gauge Mediation \vspace{18pt}}

\author{\normalsize Sibo~Zheng \vspace{12pt}\\
{\it\small   Department of Physics, Chongqing University, Chongqing 401331, P.R. China}\\}

\date{}
\maketitle \voffset -.3in \vskip 1.cm \centerline{\bf Abstract}
\vskip .3cm  After the discovery of SM-like Higgs with $m_{h}=125$ GeV,
it is increasingly urgent to explore a solution to the hierarchy problem.
In the context of MSSM from gauge-mediated SUSY breaking,
the lower bound on gluino mass suggests that the messenger scale $M$ is probably large if the magnitude of $\Lambda\sim 100$ TeV .
In this paper, we study $\overline{\mathbf{5}}+ \mathbf{5}$ model with $M\sim 10^{8}-10^{12}$ GeV and $\Lambda\simeq 100$ TeV.
For moderate Higgs-messenger coupling,
it will be shown a viable model with moderate fine tuning.
In this model,  $\mu\sim 800$ GeV and $B_{\mu}$ nearly vanishes at input scale,
which can be constructed in microscopic model.

\vskip 5.cm \noindent 08/2013
 \thispagestyle{empty}

\end{titlepage}
\section{Introduction}
A SM-like Higgs boson with mass $m_{h}=125$ GeV \cite{125} has been reported by the ATLAS and CMS collaborations at the Large Hadron Collider (LHC). Its implications in the context of minimal supersymmetric standard model (MSSM) have been extensively explored, see, e.g, \cite{MSSM1,MSSM2,MSSM3,MSSM4,MSSM5,MSSM6,MSSM7,MSSM8,MSSM9,MSSM10,MSSM11,MSSM12,MSSM13}.
Specifically, large loop-induced correction to $m_h$ is needed,
this requires either large stop mass $\sim $ several TeV for $A_{t}=0$,
 or $\sim 1$ TeV for maximal mixing.
Since it isn't promising to examine the former case at the LHC,
we focus on the later case, in which there is a large or at least moderate value of $A_{t}$ term $\sim1$ TeV.

In scenario of gauge-mediated supersymmetry (SUSY) breaking \cite{GMSB} (For a modern review, see, e.g., \cite{GMSBr}.),
$A$ term vanishes at one-loop level in models of minimal gauge mediation (GM).
It is impossible to obtain large $A_t$ in virtue of renormalization group equations (RGEs) for low-scale GM,
except that it receives non-vanishing input value at the messenger scale $M$.
According to the one-loop calculation about $A_t$ \cite{0112190, 9706540},
it requires that we must add new Yukawa couplings between the Higgs sector and the messengers in the superpotential \cite{1203.2336,1206.4086}.

However, a little hierarchy of $A-m^{2}_{H}$ similar to $\mu-B_{\mu}$ is usually induced
after adding new Yukawa couplings between the Higgs sector and the messengers,
i.e, an one-loop $A$ term accompanied with a large, two-loop, positive $m^{2}_{H}$.
Large and positive $m^{2}_{H}$ spoils the electroweak symmetry breaking (EWSB).
In order to evade this problem,
the messengers should couple to $X$ in the same way as in the minimal GM.
In order to generate one-loop $A_t$,
Yukawa couplings between (some of) messengers and doublet $H_{u}$ are added to the superpotential.
In terms of imposing additional global symmetry, the coupling of doublet $H_d$ to messengers can be forbidden.
Together with an one-loop, negative, $(\Lambda/M)^{2}$-suppressed contribution to $m^{2}_{H_{u}}$,
it can be driven to negative at electroweak (EW) scale EWSB even when $M \sim 10^{2}$ TeV.
In other words, EWSB is still viable in low-scale GM.

The arguments that leads to viable low-scale GM are obviously corrected if we consider the latest data about gluino \cite{Gluino} at the LHC.
Still we keep a low value of $\Lambda\equiv F/M\sim 100$ TeV,
with $M$ and $\sqrt{F}$ the messenger scale and SUSY-breaking scale, respectively.
It controls the whole magnitude of soft mass parameters at input scale $\sim 1$ TeV \footnote{As noted in the previous discussions, larger value of $\Lambda> 100$ TeV implies that the detection of SUSY signal at the LHC is impossible.}.
The lower bound on gluino mass $M_{3} >1$ TeV suggests $M$ must be moderate or large (up to messenger number $N$).
For $\Lambda/M \lesssim \frac{1}{4\pi}$ (or $M \gtrsim 10^{3}$ TeV) the one-loop, negative, $(\Lambda/M)^{2}$ suppressed contribution
to $m^{2}_{H_{u}}$ isn't significant.
The EWSB seems impossible for this range of $M$.

In this paper, we continue to address GM with messenger scale $M > 10^{3}$ TeV.
There are three main motivations for this study.
At first, for large messenger scale it is easy to accommodate the lower bound on gluino mass.
Second, it is also possible to achieve A term as required in terms of large RGE,
thus providing 125 GeV Higgs boson.
Finally, we will find that instead of the negative, one-loop, $(\Lambda/M)^{2}$ suppressed contribution,
the large RG corrections to $m^{2}_{H_{u}}$ due to high messenger scale take over,
thus providing EWSB.
We will show that for moderate value of Higgs-messenger coupling there indeed exists viable parameter space.

On the other hand, in contrast to SUSY models such as NMSSM,
large value of $\tan\beta$ is required in order to induce $m_{h}=125$ GeV in the MSSM.
In large $\tan\beta$ limit
the four conditions of EWSB reduce to two simple requirements:
an one-loop magnitude of $\mu$ term and vanishing $B_{\mu}$ (at least at two-loop level) at the input scale,
which can be constructed in microscopic model \cite{Zheng13}.
Alternatively, this type of $\mu-B_{\mu}$ is a consequence of adding $\mu$ term by hand,
i.e, $\mu$ term exists in SUSY limit.

The paper is organized as follows.
In section 2, we explore the model in detail.
At fist, we find the parameter space composed of $N$, $M$ and $\lambda_{u}$ (with $\Lambda=100$ TeV fixed) that gives rise to $m_{h}=125$ GeV.
Then we continue to discuss stringent constraints arising from EWSB,
which finally sets the magnitude of $\mu\sim 800$ GeV.
Put all these results together,
we argue that the model of $\overline{\mathbf{5}}+ \mathbf{5}$ that fills out $SU(5)$ is viable in GM with $M\sim 10^{8}-10^{12}$ GeV if $\alpha_{\lambda_{u}}\sim 0.02-0.04$.
Finally we conclude in section 3.

\section{The Model}
We follow the formalism of spurion superfield $X=M+\theta^{2}F$,
which stores the information of SUSY-breaking hidden sector.
The visible sector is the ordinary MSSM, which communicates with hidden $X$-sector via messenger sector.
The messengers $\phi_{i},\tilde{\phi}_{i}$ (with number of pairs $N$) fill out
either $\overline{\mathbf{10}}+ \mathbf{10}$ or $\overline{\mathbf{5}}+\mathbf{5}$ of $SU(5)$,
which can ensure grand unification still viable.
The superpotential is chosen to be,
\begin{eqnarray}{\label{spotential}}
W=X\phi_{i}\tilde{\phi}_{i}+\lambda_{uij}H_{u}\cdot \phi_{i}\cdot\tilde{\phi}_{j}
\end{eqnarray}
As noted in the introduction and argued in \cite{1206.4086},
this choice which can be realized by global symmetry reconciles the $A-m^{2}_{H_{u}}$ problem.

The total contributions to
$m^{2}_{H_{u}}$ at input scale $M$ are mainly composed of two parts:
\begin{eqnarray}{\label{mh}}
m^{2}_{H_{u}}\mid_{M}&=& \Lambda^{2}\left[2N\left(\frac{\alpha_{2}}{4\pi}\right)^{2}C_{2}(H_{u})+2N\left(\frac{\alpha_{1}}{4\pi}\right)^{2}C_{1}(H_{u})
\right.\nonumber\\
&+&\left. d_{H}(d_{H}+3)\left(\frac{\alpha_{\lambda_{u}}}{4\pi}\right)^{2}-d_{H}C_{r}\frac{\alpha_{r}\alpha_{\lambda_{u}}}{8\pi^{2}}\right]
\end{eqnarray}
with the first line in \eqref{mh} arising from the minimal GM,
and the second line in \eqref{mh} arising from the Yukawa coupling in \eqref{spotential}.
Here $C_{i}(H_{u})$ being the Casimir for $H_{u}$,
$\alpha_{r}$ being the SM gauge couplings ($r=1,2,3$),
$d_{H}$ being the effective number of messengers coupled to Higgs,
and $C_{r}=C_{H_{u}}^{r}+C_{i}^{r}+C_{j}^{r}$, $i,j$ referring to the messengers.
This same Yukawa coupling also induces deviations to $A_t$, etc.,
from that of minimal GM \footnote{For all other soft mass parameters such as sfermion masses and gaugino masses,
they are nearly the same as in the minimal GM.},
\begin{eqnarray}{\label{A}}
A_{t}&=&-d_{H}\frac{\alpha_{\lambda_{u}}}{4\pi}\Lambda,\nonumber\\
\delta m^{2}_{Q_{3}}&=&-d_{H}\frac{\alpha_{t}\alpha_{\lambda_{u}}}{16\pi^{2}}\Lambda^{2},\\
\delta m^{2}_{u_{3}}&=&-d_{H}\frac{\alpha_{t}\alpha_{\lambda_{u}}}{8\pi^{2}}\Lambda^{2}.\nonumber
\end{eqnarray}

In what follows, we consider a type of $\overline{\mathbf{5}}+ \mathbf{5}$ model
\footnote{In this paper we don't consider $\overline{\mathbf{10}}+\mathbf{10}$ in detail,
except that we will compare them with $\overline{\mathbf{5}}+ \mathbf{5}$ model somewhere in the text.},
in which the $SU(3)\times SU(2)\times U(1)$ representations of messengers take the form:
\begin{eqnarray}{\label{phi}}
(\phi_{1}, \phi_{2},\phi_{3})=((\mathbf{1},\mathbf{1},0),(\mathbf{1},\mathbf{2},1/2),(\mathbf{3},\mathbf{1},-1/3))
\end{eqnarray}
The superpotential \eqref{spotential} reads explicitly,
\begin{eqnarray}
W=X\phi_{i}\tilde{\phi}_{i}+\lambda_{u}H_{u}\cdot\phi_{1}\cdot\tilde{\phi}_{2}
\end{eqnarray}
For this model, the number of messenger pairs $N=d_{H}$.

As shown in the last line of \eqref{mh},
the modifications to $m^{2}_{H_{u}}$ due to Higgs-messenger coupling
is controlled by $d_{H}$ and the magnitude of Yukawa coupling $\lambda_{u}$.
Also note that the one-loop, negative,
$(\Lambda/M)^{2}$-suppressed contribution is tiny in comparison with those in \eqref{mh} for $\Lambda/M << 1/4\pi$.

For low-scale gauge mediation,
the appearance of one-loop, negative,$(\Lambda/M)^{2}$-suppressed contribution guarantees
that we obtain negative $m^{2}_{H_{\mu}}$ at EW scale \cite{1206.4086}.
The authors of Ref. \cite{1203.2336} focused on the possibility of driving negative $m^{2}_{H_{u}}$
by the $\alpha_{3}\alpha_{\lambda_{u}}$ term in \eqref{mh}.
In our case, as we have emphasized in the previous section,
we consider large RG running due to high messenger scale.
Therefore, the central point in our note is that the positive $m^{2}_{H_{\mu}}$ at the input scale
is driven to be negative at EW scale by large RG correction.

The parameter space is described by the following four parameters,
\begin{eqnarray}{\label{parameters}}
\left(d_{H}, ~\lambda_{u}, ~\Lambda,~ M\right)
\end{eqnarray}
The first three determine the input values of soft mass parameters at the messenger scale,
while the last one controls the magnitudes of RG corrections when we run from $M$ to EW scale.

By setting $m_{h}=125$ GeV,
we can fix one parameter, let us chose $\Lambda$.
Note that to obtain a natural EWSB it suggests $\Lambda\lesssim 100$ TeV,
while to generate a large gaugino mass which exceeds 1 TeV sets a lower bound on $N\Lambda$.
Thus, with a specific N, $\Lambda$ can be tightly constrained to be in a narrow range.

By imposing negative $m^{2}_{H_{u}}$ as favored by the EWSB,
one can constrain the magnitude of $\lambda_{u}$.
If $\lambda_{u}$ is rather small, it will induce too small $A_{t}$ term.
Conversely, if it is rather large, there will be impossible to drive the $m^{2}_{H_{u}}$ to be negative at EW scale.
With an estimate on the range of $\lambda_{u}$ in hand,
the constraint on $A_{t}$ at the messenger scale can be explicitly derived.
In virtue of RGE for $A_{t}$,
one builds the connection between the value of $A_{t}$ required by $m_{h}=125$ GeV at the EW scale
and that required by negative $m^{2}_{H_{u}}$ at the scale $M$.
This in turn determines the allowed range of $M$.

Put all these observations together,
we can examine the EWSB in large-$\tan\beta$ limit
in the parameter space of \eqref{parameters} favored by above requirements.

\subsection{Constraints from $m_{h}=125$ GeV}
The two-loop mass of Higgs boson in the MSSM reads \cite{MSSM5,9504316},
\begin{eqnarray}{\label{125}}
m_{h}^{2}&=& m_{Z}^{2}\cos^{2}2\beta+\frac{3m^{4}_{t}}{4\pi^{2}\upsilon^{2}}\left\{\log\left(\frac{M^{2}_{S}}{m^{2}_{t}}\right)
+\frac{X^{2}_{t}}{M^{2}_{S}}\left(1-\frac{X^{2}_{t}}{12M^{2}_{S}}\right)\right.\nonumber\\
&+&\left.\frac{1}{16\pi^{2}}\left(\frac{3}{2}\frac{m^{2}_{t}}{\upsilon^{2}}-32\pi \alpha_{3}\right)\left[\frac{2X^{2}_{t}}{M^{2}_{S}}\left(1-\frac{X^{2}_{t}}{12M^{2}_{S}}\right)
+\log\left(\frac{M^{2}_{S}}{m^{2}_{t}}\right)\right]\log\left(\frac{M^{2}_{S}}{m^{2}_{t}}\right)\right\}
\end{eqnarray}
where $X_{t}=A_{t}-\mu\cot\beta\simeq A_{t}$ in large-$\tan\beta$ limit,
$M_{S}=\sqrt{m_{\tilde{t}_{1}}m_{\tilde{t}_{2}}}$ being the average stop mass,
and $\upsilon=174$ GeV.

We show in fig.1 the plots of $m_{h}=125\pm 1$ GeV as function of $A_{t}$
with fixed $\Lambda= 10^{2}$ TeV for $M=10^{8}$ GeV.
The three colors represent different numbers of messenger pairs
\footnote{One can examine that for $\Lambda=10^{2}$ TeV,
the parameter space $N\geqslant1$ and $M\sim 10^{8}-10^{12}$ GeV can give rise to RGE large enough for $M_{3}$  such that at EW scale its value is larger than  mass bound $\sim$ 1 TeV \cite{Gluino} reported by the LHC experiments.}.
Fig.1 shows that for $N=2,3$ $\mid A_{t}\mid \sim 1.0-1.5$ TeV at EW scale can provide 125 GeV.
As $M$ increases to $\sim 10^{12}$ GeV,
there is no significant  modification to the value of  $A_{t}$ as required.
\begin{figure}[h!]
\centering
\centering
\begin{minipage}[b]{0.8\textwidth}
\centering
\includegraphics[width=4.5in]{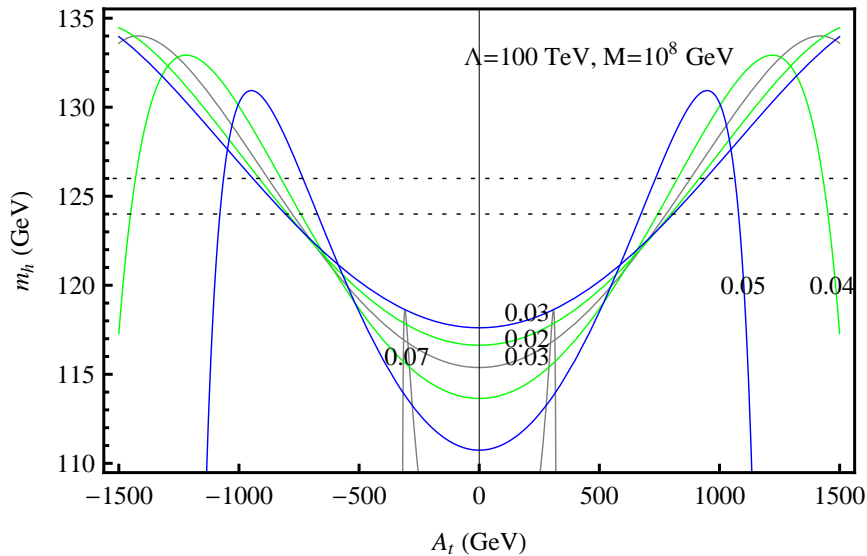}
\end{minipage}%
\caption{Plots $m_{h}=125\pm 1$ GeV as function of $A_{t}$ for $\Lambda= 10^{2}$ TeV , $M=10^{8}$ GeV and different values of $N$ and $\alpha_{\lambda_{u}}$.
The value of $A_{t}$ is shown at EW other than input scale.
The gray, green and blue represent  $N=1$, $N=2$ and $N=3$, respectively.
For each case of $N$, the values of $\alpha_{\lambda_{u}}$s are explicitly shown in the plots.
The dotted horizontal lines correspond to the range of $124-126$ GeV.}
\end{figure}

\begin{figure}[h!]
\centering
\begin{minipage}[b]{0.8\textwidth}
\centering
\includegraphics[width=4.5in]{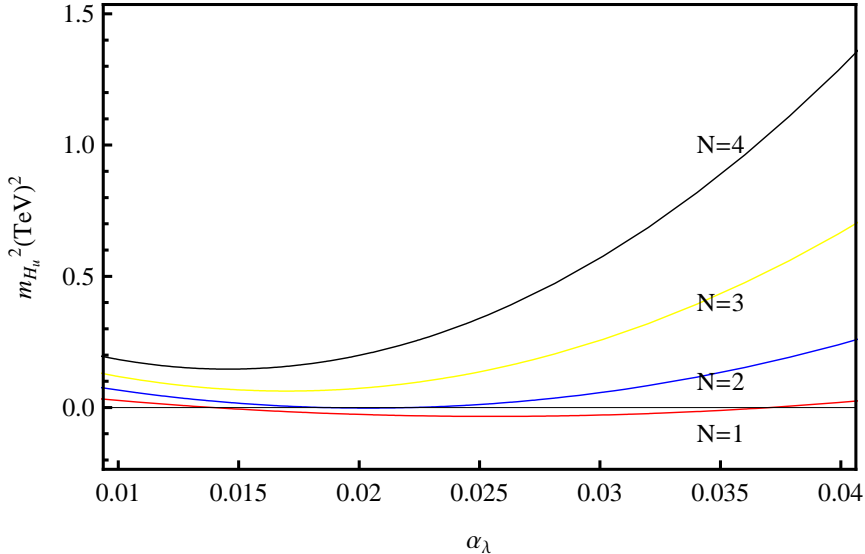}
\end{minipage}%
\caption{Plots of $m^{2}_{H_{u}}$ at the input scale as function of $\lambda_{u}$ with different $N$s for $\Lambda=100$ TeV and $M=10^{8}$ GeV.
Negative $m^{2}_{H_{u}}$ can be obtained in the range $0.02 \lesssim\alpha_{\lambda_{u}}\lesssim 0.04$ for $N=1$.
Increasing $N$ the range allowed will shrink.
In particular, only the neighborhood of $\alpha_{\lambda_{u}}\simeq0.02$ is possible for $N=2$. Similar plots can be found for different choices of $M$.}
\end{figure}

We show in fig.2 plots of $m^{2}_{H_{u}}$ at input scale
as function of $\lambda_{u}$ with different $N$s for $\Lambda=100$ TeV and $M=10^{8}$ GeV.
Negative $m^{2}_{H_{u}}$ can be obtained in the range $0.02 \lesssim\alpha_{\lambda_{u}}\lesssim 0.04$ for $N=1$.
Increasing $N$ the range allowed will shrink.
In particular, only the neighborhood of $\alpha_{\lambda_{u}}\simeq0.02$ is possible for $N=2$.
Similar plots can be found for $M=10^{12}$ GeV.
Following fig.2 one observes that the region $\alpha_{\lambda_{u}}> 0.1$ induces positive $m^{2}_{H_{u}}$  too large at input scale to be driven negative at EW scale.
Actually, the positivity of stop soft masses at input scale require $\alpha_{\lambda_{u}}< 0.1$.
On the other hand,  the region $\lambda_{u}< 0.01$ provides $\mid A_{t}\mid $ term
too small at input scale to accommodate required value of $\mid A_{t}\mid \gtrsim 1$ TeV at EW scale.
Fig. 1 and 2 show us the possible range for $\alpha_{\lambda_{u}}$ as
\begin{eqnarray}{\label{lambda}}
0.01\lesssim \alpha_{\lambda_{u}}< 0.1
\end{eqnarray}
Corresponding the range of $\mid A_t\mid$ is from several hundred GeV to $\sim$ 1 TeV at the input scale.
Similar plots can also be found for $\overline{\mathbf{10}}+ \mathbf{10}$.
For realistic EWSB, fig.2 shows that large RGE for $m^{2}_{H_{u}}$ must take over in the most range of \eqref{lambda}.

Now we consider the connection between the values of $A_{t}$ at input and EW scale in virtue of fig.1.
Roughly we need a correct RGE which ensures that $A_{t}$ runs from a correct value at messenger scale to $\sim -1$ TeV at the EW scale.
Recall the beta function $\beta_{A_{t}}$ below the messenger scale for $A_{t}$ \cite{9709356},
\begin{eqnarray}{\label{beta}}
\beta_{A_{t}}=\frac{1}{4\pi}\left[18\alpha_{t}A_{t}-\frac{16}{3}\alpha_{3}(A_{t}-2y_{t}M_{3})
-3\alpha_{2}(A_{t}-2y_{t}M_{2})-\frac{13}{15}\alpha_{1}(A_{t}-2y_{t}M_{1})\right]\nonumber\\
\end{eqnarray}
Since the sign of $\beta_{A_{t}}$ depends on the relative magnitude of $\mid A_{t}\mid$ to $M_{3}$,
or concretely the magnitude of $\alpha_{\lambda_{u}}$ to $\alpha_{3}$,
it can be either positive or negative in of the range of \eqref{lambda}.
\begin{figure}[h!]
\centering
\begin{minipage}[b]{0.8\textwidth}
\centering
\includegraphics[width=4in]{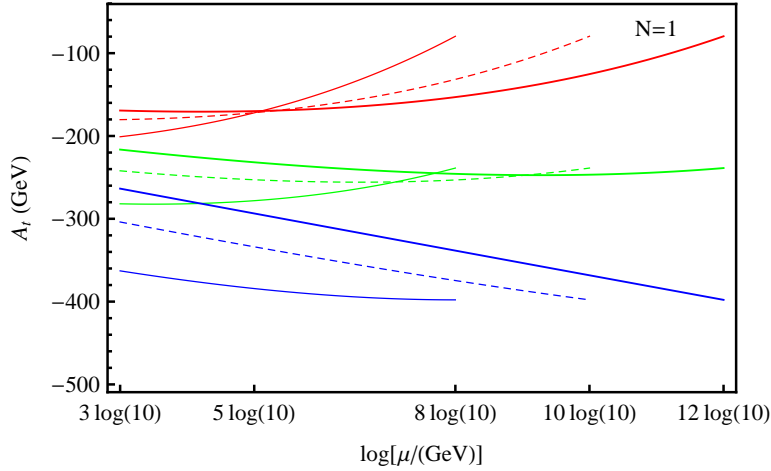}
\end{minipage}%
\caption{RGEs for $A_t$ as function of $\alpha_{\lambda_{u}}$ and input messenger scale $M$ in the case $N=1$ and $\Lambda=100$ TeV.
The red, green and blue correspond to $\alpha_{\lambda_{u}}=0.01$, $0.03$, and $0.05$, respectively.
The solid, dashing and thickness refer to $M=10^{8}$, $10^{10}$ and $10^{12}$ (GeV), respectively.  }
\end{figure}
\begin{figure}[h!]
\centering
\begin{minipage}[b]{0.8\textwidth}
\centering
\includegraphics[width=4in]{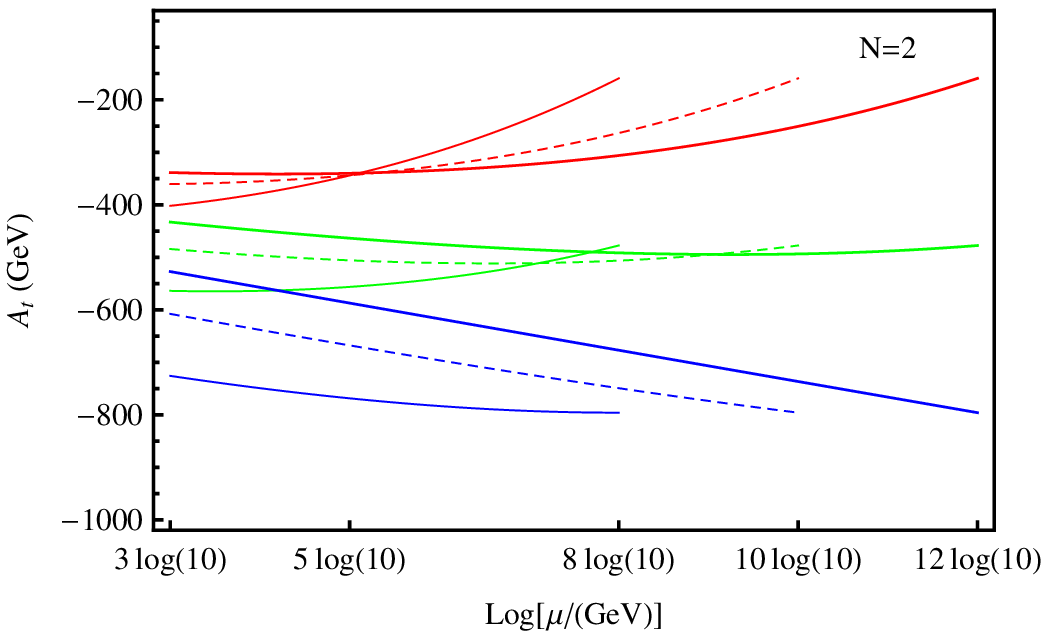}
\end{minipage}%
\caption{Same as Fig.3 for $N=2$.}
\end{figure}
\begin{figure}[h!]
\centering
\begin{minipage}[b]{0.8\textwidth}
\centering
\includegraphics[width=4in]{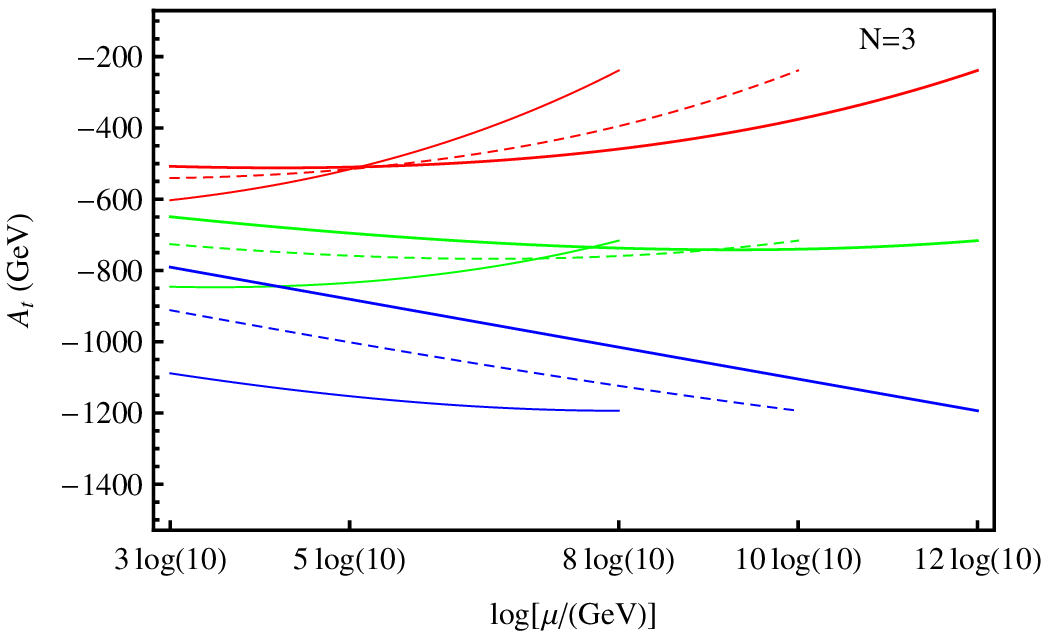}
\end{minipage}%
\caption{Same as Fig.3 for $N=3$.}
\end{figure}

By using \eqref{beta} we show the RGE for $A_{t}$ in the cases $N=1$ (fig.3), $N=2$ (fig.4) and $N=3$ (fig.5),
from which the dependence on $\alpha_{\lambda_{u}}$ and $M$ is obvious.
In particular $N=1$ is excluded (as shown from fig. 3) regardless of the values of $M$ and $\alpha_{\lambda_{u}}$.
And $\alpha_{\lambda_{u}}<0.05$ is excluded for the case $N=2$.
In the case of $N=3$, the parameter pace allowed is possible for the whole range of $M\sim 10^{8}-10^{12}$ GeV
and for $\alpha_{\lambda_{u}}\sim 0.03-0.05$
In summary, the parameter space for $\Lambda=10^{2}$ TeV which can explain the
125 GeV Higgs boson is restricted to region $N>2$ and $\alpha_{\lambda_{u}}\sim 0.03-0.05$.
In the next subsection, we will explore EWSB in this narrow region,
determine soft breaking masses at EW scale, and measure the fine tuning in this type of  model.

\subsection{EWSB}
In the previous section we address the parameter space which provides $m_{h}=125$ GeV at the EW scale.
In what follows, we continue to explore another question whether this parameter space induces the EWSB simultaneously.
We begin with the conditions of EWSB involving soft parameters in the Higgs sector in the MSSM:
\begin{eqnarray}{\label{ewsb1}}
\mu^{2}&=&\frac{m^{2}_{H_{d}}-m^{2}_{H_{u}}\tan^{2}\beta}{\tan^{2}\beta-1}-\frac{m^{2}_{Z}}{2}\nonumber\\
\sin 2\beta &=&\frac{2B_{\mu}}{m^{2}_{H_{d}}+m^{2}_{H_{u}}+2\mu^{2}}
\end{eqnarray}
which together with
\begin{eqnarray}{\label{ewsb2}}
B_{\mu}&<&\frac{1}{2}(m^{2}_{H_{d}}+m^{2}_{H_{u}})+\mid\mu\mid^{2}\nonumber\\
B^{2}_{\mu} &>&(\mid\mu\mid^{2}+m^{2}_{H_{d}})(\mid\mu\mid^{2}+m^{2}_{H_{u}}).
\end{eqnarray}
guarantee a stable vacuum.
In the large $\tan\beta$ limit together with negative $m^{2}_{H_{u}}$ and positive $m^{2}_{H_{d}}$,
\eqref{ewsb1} and  \eqref{ewsb2} reduce to,
\begin{eqnarray}{\label{ewsb}}
\mu^{2}&=&-m^{2}_{H_{u}}-\frac{m^{2}_{Z}}{2}\nonumber\\
B_{\mu}&<<&\frac{1}{2}(m^{2}_{H_{d}}+m^{2}_{H_{u}})+\mid\mu\mid^{2}=\frac{1}{2}(m^{2}_{H_{d}}-m^{2}_{H_{u}}-m^{2}_{Z})
\end{eqnarray}
The first constraint in \eqref{ewsb} fixes the magnitude of $\mu$ at the input scale $\mu(M)$ through RGE for $\mu$,
and the last of which is the new constraint to be satisfied.
\begin{figure}[h!]
\centering
\begin{minipage}[b]{0.8\textwidth}
\centering
\includegraphics[width=4in]{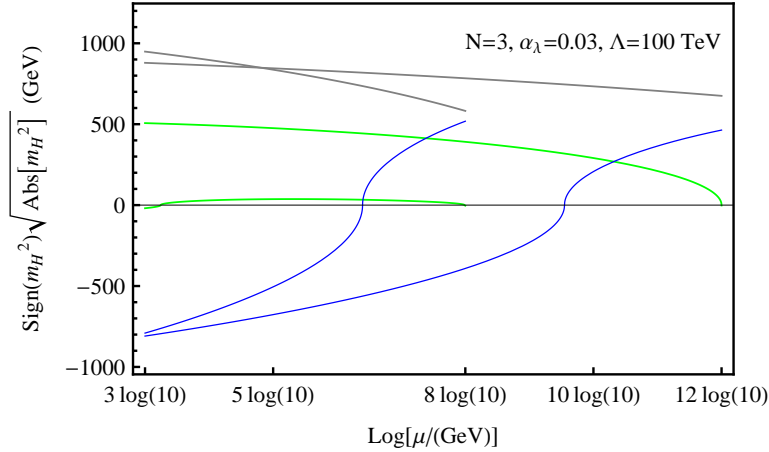}
\end{minipage}%
\caption{RGEs for $Sign[m^{2}_{i}]\sqrt{\mid m^{2}_{i}\mid }$ with $M=10^{8}, 10^{12}$ GeV and $\alpha_{\lambda_{u}}=0.03$.
The gray, blue and green curves refer to $m^{2}_{H_{d}}$, $m^{2}_{H_{u}}$ and $B_{\mu}$, respectively.  }
\end{figure}

In traditional MSSM without Yukawa coupling $\lambda_u$,
$m^{2}_{H_{u}}$ is driven negative at EW scale through RGE.
In our model, similar phenomenon appears for small $\alpha_{\lambda_{u}}$.
Otherwise, the the input values significantly increase for $m^{2}_{H_{u}}$ and decrease for stop masses squared due to large $\alpha_{\lambda_{u}}$,
which would spoil EWSB.
In fig.6 we show RGEs for $m_{H_{u}}$ and $m_{H_{d}}$ for input scale $M=10^{8}$ GeV ($10^{12}$ GeV) and
$\alpha_{\lambda_{u}}=0.03$  for the case $N=3$.
Larger value of $\alpha_{\lambda_{u}}$ isn't viable.
Following fig. 6 we obtain $\mu\sim 800 $ GeV at EW scale.
With this value of $\mu$ term, we can estimate the REG for $B_{\mu}$ as follows.
In our model, $B_{\mu}\simeq 0$ (at two-loop level) at the input scale,
therefore the second condition of \eqref{ewsb} can be trivially satisfied in virtue of REG for $B_{\mu}$,
\begin{eqnarray}{\label{bmu}}
16\pi^{2}\beta_{B_{\mu}}\simeq B_{\mu}\left(3y^{2}_{t}-3g^{2}_{2}-\frac{3}{5}g^{2}_{1}\right)+\mu\left(6y_{t}A_{t}+6g^{2}_{2}M_{2}+\frac{6g^{2}_{1}}{5}M_{1}\right)
\end{eqnarray}
For $M=10^{8}(10^{12})$ GeV substituting $\mu\sim 800$ GeV (from fig.6) into \eqref{bmu}
and using the REGs for $A_{t}$ (fig.5) and $M_{1,2}$,
one obtains the RGE for $B_{\mu}$ term in each case \footnote{It is crucial to note that the sign of $B_{\mu}$ must be positive,
even if its absolute value is rather small in comparison with the magnitude of $(m^{2}_{H_{d}}-m^{2}_{H_{u}})$.},
as shown by the green plots in fig.6.
One finds that for small input scale $M=10^{8}$ GeV,
$B_{\mu}\sim (10~GeV)^{2}$.
As we adjust the value of $M$ to $10^{12}$ GeV,
the RG correction to $B_{\mu}$ increases to $(500~GeV)^{2}$ .

Finally, we use the traditional definition $c=\max\{c_{i}\}$ to measure fine tuning in the model,
where $c_{i}=\partial \ln m^{2}_{Z}/\partial \ln m^{2}_{i}$, $m^{2}_{i}$ being soft breaking mass squared.
For mass spectrum in fig.6,  the main contribution to large value of $c$ arises from stop masses, gaugino masses and $\mu$ term.
Typically we have $c\simeq150-200$,
which suggests that this type of model is moderately fine tuned.

\section{Conclusions}
The discovery of SM-like Higgs with $m_{h}=125$ GeV verifies SM as a precise low-energy effective theory.
It is increasingly urgent to find a solution to stabilize the mass of this scalar.
At present status, SUSY is still in the short list of frameworks which can provide such a solution with some fine tunings.
This paper is devoted to explore gauge-mediated SUSY with latest results reported by the LHC experiments.

The constraint of $m_{h}=125$ GeV and the lower bound on $\Lambda$ needs an $A_t$ term $\sim 1$ TeV.
However, since the vanishing of $A_t$ soft term at input scale at one-loop level,
it is impossible to obtain such large value in low-scale GM
except we either introduce direct Higgs-messenger coupling or consider high messenger scale.

The situation is rather different in GM with high messenger scale (with $\Lambda\sim 100$ TeV fixed).
At first, the negative $m^{2}_{H_{u}}$ required by EWSB must be realized due to large RG correction instead of the one-loop,
negative, $(\Lambda/M)^{2}$-suppressed contribution \cite{1206.4086}.
Second, the $A_t$ term can still be obtained with small or moderate Higgs-messenger coupling.
Finally, the large lower bound on gluino mass suggests that large messenger scale is favored.
Therefore, it is necessary to explore viable GM with $M > 10^{7}$ GeV.
Following these motivations,
we find that a type of $\overline{\mathbf{5}}+ \mathbf{5}$ model \cite{1206.4086} is viable in GM with $M\sim 10^{8}-10^{12}$ GeV
and moderate value of Higgs-messenger coupling $\alpha_{\lambda_{u}}$.
In this model, $m^{2}_{H_{u}}$ is driven to be negative although it has positive input value $\sim \mathcal{O}(1)$ TeV$^{2}$.
At EW scale we obtain EWSB with the magnitude of $\mu\sim 800$ GeV,
$m_{h}=125$ GeV and $M_{3}>$ 1 TeV.

This magnitude of $\mu$ term can be either generated at one-loop level at input scale,
with $B_{\mu}$ vanishing at least at two-loop level,
or considered as an input scale in SUSY limit.
For the first case, we refer the reader to \cite{Zheng13},
where this type of $\mu$ and $B_{\mu}$ terms can indeed be realized in terms of adding SM singlets to the MSSM.
The latter choice makes the model complete,
although it seems ad hoc to add a $\mu$ term by hand.  \\

~~~~~~~~~~~~~~~~~~~~~~~~~~~~~~~~~~~~~~~~
$\bf{Acknowledgement}$\\
This work is supported in part by the National Natural Science Foundation of China with grant No.11247031.\\

\end{document}